\documentclass[aps,twocolumn,preprintnumbers,superscriptaddress,showpacs,nofootinbib,amsmath, amssymb]{revtex4-2}

\usepackage{graphicx}
\usepackage{subfigure}
\usepackage{hyperref}
\usepackage{color}
\usepackage[table]{xcolor}
\usepackage{boldline,multirow}
\usepackage{booktabs}
\usepackage{pifont}
\usepackage{mathrsfs}
\usepackage{soul}
\usepackage[utf8]{inputenc}
\usepackage{multirow}
\usepackage{atbegshi,picture}

\usepackage{color}
\definecolor{rosso}{cmyk}{0,1,1,0.4}
\definecolor{rossos}{cmyk}{0,1,1,0.55}
\definecolor{rossoc}{cmyk}{0,1,1,0.2}
\definecolor{blu}{cmyk}{1,1,0,0.3}
\definecolor{blus}{cmyk}{1,1,0,0.6}
\definecolor{bluc}{cmyk}{1,1,0,0.1}
\definecolor{verde}{cmyk}{0.92,0,0.59,0.25}
\definecolor{verdec}{cmyk}{0.92,0,0.59,0.15}
\definecolor{verdes}{cmyk}{0.92,0,0.59,0.4}
\definecolor{bviolet}{rgb}{0.54, 0.17, 0.89}
\definecolor{myred}{rgb}{0.545,0.004,0}
\hypersetup{colorlinks,bookmarksopen,bookmarksnumbered,citecolor=myred,
linkcolor=myred,pdfstartview=FitH,urlcolor=myred}

\newcommand{\beq}{\begin{equation}} 
\newcommand{\eeq}{\end{equation}}
\newcommand{\bea}{\begin{eqnarray}}  
\newcommand{\eea}{\end{eqnarray}}
\newcommand{\beastar}{\begin{eqnarray*}}  
\newcommand{\eeastar}{\end{eqnarray*}}
\newcommand{\nnl}{\nonumber \\}

\definecolor{Gray}{gray}{0.9}
\newcommand{\cg}{\cellcolor{Gray}}

\AtBeginShipoutNext{\AtBeginShipoutUpperLeft{%
  \put(\dimexpr\paperwidth-10cm\relax,-1cm){\makebox[1pt][r]{In loving memory of Enrico Franco, Scientist, Mentor and Friend.}}%
}}

\begin{document}

\preprint{P3H-22-127}
\preprint{TTP22-073}
\preprint{YITP-SB-22-42}

\title{Constraints on Lepton Universality Violation from Rare \texorpdfstring{\boldmath$B$}{B} Decays}

\author{Marco Ciuchini}
\email[]{marco.ciuchini@roma3.infn.it}
\affiliation{INFN Sezione di Roma Tre,
Via della Vasca Navale 84, I-00146 Rome, Italy}

\author{Marco Fedele}
\email[]{marco.fedele@kit.edu}
\affiliation{Institut f\"ur Theoretische Teilchenphysik, Karlsruhe Institute of Technology, D-76131 Karlsruhe, Germany}

\author{Enrico Franco}
\email[]{enrico.franco@roma1.infn.it}
\affiliation{INFN Sezione di Roma, Piazzale Aldo Moro 2, I-00185 Rome, Italy}

\author{Ayan Paul}
\email[]{a.paul@northeastern.edu}
\affiliation{Electrical and Computer Engineering, Northeastern University, Boston MA 02115, USA}

\author{Luca Silvestrini}
\email[]{luca.silvestrini@roma1.infn.it}
\affiliation{INFN Sezione di Roma, Piazzale Aldo Moro 2, I-00185 Rome, Italy}

\author{Mauro Valli}
\email[]{mauro.valli@roma1.infn.it}
\affiliation{INFN Sezione di Roma, Piazzale Aldo Moro 2, I-00185 Rome, Italy}
\affiliation{C.N. Yang Institute for Theoretical Physics, Stony Brook University, Stony Brook, NY 11794,~USA}

\begin{abstract}
 The LHCb collaboration has very recently released a new study of $B^+ \to K^{+} \ell^+ \ell^-$ and $B \to K^{*0} \ell^+ \ell^-$ ($\ell = e,\mu$) decays, testing lepton universality with unprecedented accuracy using the whole Run 1 and 2 dataset. In addition, the CMS collaboration has recently reported an improved analysis of the branching ratios $B_{(d,s)}\to\mu^+\mu^-$. While these measurements offer, per se, a powerful probe of New Physics, global analyses of $b \to s \ell^+ \ell^-$ transitions also rely on the assumptions about nonperturbative contributions to the decay matrix elements. In this work, we perform a global Bayesian analysis of New Physics in (semi)leptonic rare $B$ decays, paying attention to the role of charming penguins which are difficult to evaluate from first principles. We find data to be consistent with the Standard Model once rescattering from intermediate hadronic states is included. Consequently, we derive stringent bounds on lepton universality violation in $|\Delta B | = | \Delta S| = 1$ (semi)leptonic processes.
\end{abstract}

\maketitle

Since the first collisions in 2010, the Large Hadron Collider (LHC) allowed for a tremendous step forward in the electroweak (EW) sector of the Standard Model (SM) of Particle Physics -- culminated with the discovery of the Higgs boson~\cite{Aad:2012tfa,Chatrchyan:2012ufa} -- while it has also excited the community with a few interesting hints of Physics Beyond the SM (BSM). In particular, the LHCb collaboration provided the first statistically relevant hint for Lepton Universality Violation (LUV) in flavor-changing neutral-current (FCNC) processes~\cite{LHCb:2014vgu}, measuring the ratio $R_{K} \equiv Br(B^{+} \to K^{+} \mu^{+} \mu^{-})/Br(B^{+} \to K^{+} e^{+} e^{-})$ in the dilepton invariant-mass range $q^{2} \in[1,6]$ GeV$^2$. These hints have been confirmed by subsequent measurements, always by the LHCb collaboration, namely $R_{K}$~\cite{LHCb:2017avl}, $R_{K^*}$~\cite{LHCb:2019hip,LHCb:2021trn}, $R_{K_S}$ and $R_{K^{*+}}$~\cite{LHCb:2021lvy}. 

Interestingly enough, these hints of LUV appeared in transitions where deviations from the SM were already claimed, see e.g.~\cite{Descotes-Genon:2013wba,Altmannshofer:2013foa,Beaujean:2013soa,Hurth:2013ssa}, on the basis of the measurements of angular distributions in $b \to s \mu^+ \mu^-$ decays \cite{LHCb:2013zuf,LHCb:2013tgx,LHCb:2013ghj,CMS:2013mkz,LHCb:2014cxe,LHCb:2015wdu,LHCb:2015svh,LHCb:2020lmf,CMS:2020oqb,LHCb:2020gog,LHCb:2021zwz,LHCb:2021xxq}. Claiming discrepancies from SM predictions in Branching Ratios (BRs) and angular distributions requires, however, full theoretical control on hadronic uncertainties in the matrix element calculation~\cite{Jager:2012uw,Descotes-Genon:2014uoa,Jager:2014rwa}, and in particular on the so-called \textit{charming penguins}~\cite{Ciuchini:1997hb}, which might affect the vector coupling to the leptons even in regions of the dilepton invariant mass well below the charmonium threshold~\cite{Lyon:2014hpa,Melikhov:2022wct} and bring the SM in agreement with experiment~\cite{Ciuchini:2015qxb}. Combining angular distributions with LUV data strengthened the case for New Physics (NP), since a single NP contribution could reproduce the whole set of data~\cite{Hiller:2021pul,Geng:2021nhg,Cornella:2021sby,Hurth:2021nsi,Alguero:2021anc,Bause:2021cna,Altmannshofer:2021qrr,Ciuchini:2021smi,Alguero:2022wkd}. On the other hand, charming penguins might affect the picture of NP behind LUV, since LUV ratios depend on the interplay of NP and hadronic contributions \cite{Ciuchini:2017mik,Ciuchini:2019usw,Ciuchini:2020gvn,Ciuchini:2021smi}. 
While considerable progress has been made in estimating (at least part of) the charming-penguin amplitudes using light-cone sum rules \cite{Khodjamirian:2010vf,Khodjamirian:2012rm} and analyticity supplemented with perturbative QCD in the Euclidean $q^2$ region \cite{Bobeth:2017vxj,Chrzaszcz:2018yza,Gubernari:2020eft,Gubernari:2022hxn}, calculating these hadronic contributions remains an open problem, as we discuss below.  

\begin{figure*}[!t!]
\centering
\includegraphics[width=\textwidth]{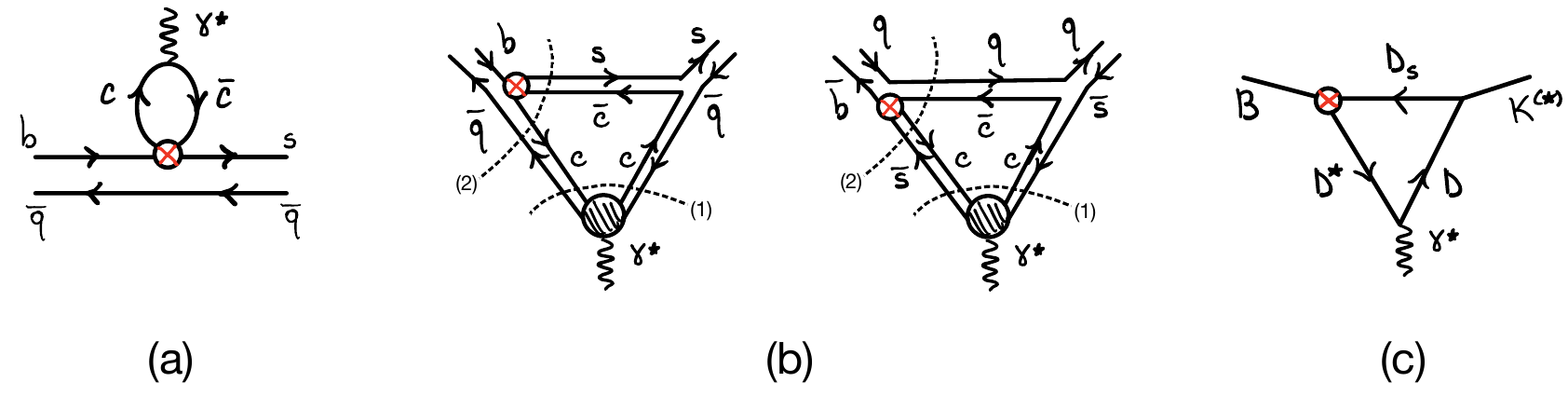}
\caption{\textit{Example of charming-penguin diagrams contributing to the $ B \to K^{(*)} \ell^+ \ell^-$ amplitude. Diagram \textbf{(a)} represents the class of charming-penguin amplitudes related to $c-\bar{c}$ state that subsequently goes into a virtual photon, see refs.~\cite{Khodjamirian:2010vf,Bobeth:2017vxj,Chrzaszcz:2018yza,Gubernari:2020eft,Gubernari:2022hxn}. Diagram \textbf{(b)}  and \textbf{(c)}  represent the kind of contributions from rescattering of intermediate hadronic states, at the quark and meson level respectively.  The phenomenological relevance of rescattering for the SM prediction of the $ B \to K^{(*)} \ell^+ \ell^-$ decays has been recently considered in ref.~\cite{Ciuchini:2021smi}.}}
\label{fig:rescattering}
\end{figure*}

Before presenting our results, we notice that very recently the experimental picture 
drawn so far has suddenly changed. 
Firstly, the CMS collaboration provided a new analysis of BR$(B_{(d,s)} \to \mu^+ \mu^-)$ with the full Run 2 dataset~\cite{CMS-PAS-BPH-21-006}, bringing the HFLAV average
\begin{equation}
    \mathrm{BR}(B_{s} \to \mu^+ \mu^-) = (3.45 \pm 0.29) \cdot 10^{-9}
    \label{eq:newBsmmExp}
\end{equation}
into excellent agreement with the SM prediction $\mathrm{BR}(B_{s} \to \mu^+ \mu^-) = (3.47 \pm 0.14) \cdot 10^{-9}$~\cite{Bobeth:2013uxa,UTfit:2022hsi}. Being short-distance dominated, this FCNC process strongly constrains NP contributions involving, in particular, axial leptonic couplings~\cite{Alonso:2014csa,Fleischer:2017ltw}. Furthermore, an updated LHCb analysis of $R_K$ and $R_{K^*}$ based on the full Run 1 and 2 dataset has been presented~\cite{LHCb:2022qnv,LHCb:2022zyk}:
\begin{eqnarray}
    \label{eq:RK22}
    R_{K_{[0.1,1.1]}} &= 0.994\;^{+0.090}_{-0.082}\;({\rm stat})\;^{+0.029}_{-0.027}\;({\rm syst})\,, \\  \nonumber
    R_{{K^{*}}_{[0.1,1.1]}} &= 0.927\;^{+0.093}_{-0.087}\;({\rm stat})\;^{+0.036}_{-0.035}\;({\rm syst})\,,\\ \nonumber 
    R_{K_{[1.1,6]}} &= 0.949\;^{+0.042}_{-0.041}\;({\rm stat})\;^{+0.022}_{-0.022}\;({\rm syst})\,, \\   \nonumber 
    R_{{K^{*}}_{[1.1,6]}} &= 1.027\;^{+0.072}_{-0.068}\;({\rm stat})\;^{+0.027}_{-0.026}\;({\rm syst})\,, 
\end{eqnarray}
with correlations reported in Fig. 26 of ref.~\cite{LHCb:2022zyk}. These new measurements
dramatically change the scenario of possible LUV effects in FCNC $B$ decays~\cite{Crivellin:2021sff}, questioning what in the last years served as fertile ground for model building, see for instance~\cite{Buras:2013qja,Buras:2014yna,Altmannshofer:2014cfa,Glashow:2014iga,Gripaios:2014tna,Barbieri:2016las,Assad:2017iib,DiLuzio:2017vat,Calibbi:2017qbu,Bordone:2017bld,Barbieri:2017tuq,Greljo:2018tuh,Marzocca:2018wcf,DiLuzio:2018zxy,Blanke:2018sro,Fornal:2018dqn,Arnan:2019uhr,Fuentes-Martin:2020bnh,Gherardi:2020qhc,Arcadi:2021cwg,Darme:2021qzw,Greljo:2021npi,Davighi:2022fer,Davighi:2022qgb,Fuentes-Martin:2022xnb}.

In this \textit{Letter} we provide a reassessment of NP effects in $b \to s\, \mu^+ \mu^-$ transitions in view of the experimental novelties discussed above. Adopting the model-independent language of the Standard Model Effective Theory (SMEFT)~\cite{Buchmuller:1985jz,Grzadkowski:2010es}, we present an updated analysis of $|\Delta B | = | \Delta S| = 1$ (semi)leptonic processes and show that current data no longer provide strong hints for NP. Indeed, updating the list of observables considered in our previous global analysis~\cite{Ciuchini:2021smi} with the results in eqs.~\eqref{eq:newBsmmExp} and~\eqref{eq:RK22}, the only remaining measurements deviating from SM expectations and not affected by hadronic uncertainties are the LUV ratios $R_{K_S}$ and $R_{K^{*+}}$~\cite{LHCb:2021lvy}, for which a re-analysis by the LHCb collaboration is mandatory in view of what discussed in~\cite{LHCb:2022qnv,LHCb:2022zyk}.

The anatomy of the $B \to K^{(*)}\ell^+\ell^-$ decay can be characterized in terms of helicity amplitudes~\cite{Jager:2012uw,Gratrex:2015hna}, that in the SM at a scale close to the bottom quark mass $m_b$ can be written as:
\bea 
H_V^{\lambda} &\propto& \left\{C_9^{\rm SM}\widetilde{V}_{L\lambda} + \frac{m_B^2}{q^2} \left[\frac{2m_b}{m_B}C_7^{\rm SM}\widetilde{T}_{L\lambda}  - 16\pi^2h_{\lambda} \right]\right\}\,,\nonumber\\
H_A^{\lambda} &\propto& C_{10}^{\rm SM}\widetilde{V}_{L\lambda} \ , \
H_P \propto \frac{m_{\ell} \, m_{b}}{q^2} \,  C_{\rm 10}^{\rm SM} \left( \widetilde{S}_{L} - \frac{m_s}{m_b}\widetilde{S}_{R} \right)\,,\nonumber
\eea
with $\lambda=0,\pm$ and $C_{7,9,10}^{\rm SM}$ the SM Wilson coefficients of the semileptonic operators of the $|\Delta B| = | \Delta S| =1 $ weak effective Hamiltonian~\cite{Buchalla:1995vs,Grinstein:2015nya,Silvestrini:2019sey}, normalized as
in ref.~\cite{Ciuchini:2019usw}. 
The naively factorizable contributions to the above amplitudes can be expressed in terms of seven $q^{2}$-dependent form 
factors, $\widetilde{V}_{0,\pm}$, $\widetilde{T}_{0,\pm}$ and $\widetilde{S}$~\cite{Straub:2015ica,Gubernari:2018wyi}. At the loop level, non-local effects parametrically not suppressed (neither by small Wilson coefficients nor by small CKM factors) arise from the insertion of the following four-quark operator:
\bea
\label{eq:quark_O}
Q^c_2 &=& (\bar{s}_L\gamma_{\mu} c_L)(\bar{c}_L\gamma^{\mu}b_L)\,,
\eea
that yields non-factorizable power corrections in $H_V^{\lambda}$ via the hadronic correlator $h_\lambda(q^2)$~\cite{Jager:2014rwa,Ciuchini:2015qxb,Chobanova:2017ghn}, receiving the main contribution from the time-ordered product:
\beq \label{eq:hlambda}
\frac{\epsilon^*_\mu(\lambda)}{m_B^2} \int d^4x\ e^{iqx} \langle \bar K^* \vert \mathcal{T}\{j^\mu_\mathrm{em}(x)
Q^{c}_{2} (0)\} \vert \bar B \rangle \,,
\eeq
with $j^\mu_\mathrm{em}(x)$ the electromagnetic (quark) current.

This correlator receives two kinds of contributions. The first corresponds to diagrams of the form of diagram \textbf{(a)} in Fig.~\ref{fig:rescattering}, where the initial $B$ meson decays to the $K^{(*)}$ plus a $c\bar{c}$ state that subsequently goes into a virtual photon. This contribution has been studied in detail in the context of light-cone sum rules in the regime $q^2 \ll 4m_c^2$ in~\cite{Khodjamirian:2010vf}; in the same reference, dispersion relations were used to extend the result to larger values of the dilepton invariant mass. While the operator product expansion performed in ref.~\cite{Khodjamirian:2010vf} was criticized in ref.~\cite{Melikhov:2022wct}, and multiple soft-gluon emission may represent an obstacle for the correct evaluation of this class of hadronic contributions~\cite{Ciuchini:2015qxb,Ciuchini:2016weo,Ciuchini:2017mik,Ciuchini:2018anp}, refs.~\cite{Bobeth:2017vxj,Chrzaszcz:2018yza} have exploited analyticity in a more refined way than~\cite{Khodjamirian:2010vf}. In those works the negative $q^2$ region -- where perturbative QCD is supposed to be valid -- has been used to further constrain the amplitude. Building on these works, together with unitarity bounds~\cite{Gubernari:2020eft}, ref.~\cite{Gubernari:2022hxn} found a very small effect in the large-recoil region.  

The second kind of contribution to the correlator in eq.~\eqref{eq:hlambda} originates from the triangle diagrams depicted in Fig.~\ref{fig:rescattering} \textbf{(b)}, in which the photon can be attached both to the quark and antiquark lines and we have not drawn explicitly the gluons exchanged between quark-antiquark pairs. An example of an explicit hadronic contribution of this kind is depicted in Fig.~\ref{fig:rescattering} \textbf{(c)}.\footnote{See ref.~\cite{Ladisa:2022vmh} for a very recent estimate of similar diagrams with up quarks, rather than charm quarks, in the internal loop.} The $D_sD^*$ pair is produced by the weak decay of the initial $B$ meson with low momentum, so that no color transparency argument holds and rescattering can easily take place. Furthermore, the recent observation of tetraquark states in $e^+e^- \to  K(D_sD^{*}+D_s^{*}D)$ by the BESIII collaboration \cite{BESIII:2020qkh} confirms the presence of nontrivial nonperturbative dynamics of the intermediate state. 

One could think of applying dispersive methods also to this kind of contributions, but the analytic structure of triangle diagrams is quite involved, depending on the values of external momenta and internal masses. A dispersion relation in $q^2$ of the kind used in refs.~\cite{Khodjamirian:2010vf,Bobeth:2017vxj,Chrzaszcz:2018yza,Gubernari:2020eft,Gubernari:2022hxn}, based on the cut denoted by (1) in Fig.~\ref{fig:rescattering} \textbf{(b)}, could be written if the $B$ invariant mass were below the threshold for the production of charmed intermediate states. However, when the $B$ invariant mass raises above the threshold for cut (2), an additional singularity moves into the $q^2$ integration domain, requiring a nontrivial deformation of the path (see for example the detailed discussion in ref.~\cite{Fronsdal}). Another possibility would be to get an order-of-magnitude estimate of contributions as the one in Fig.~\ref{fig:rescattering} \textbf{(c)} using an approach similar to ref.~\cite{Ladisa:2022vmh}. 

To be conservative, and in the absence of a first-principle calculation of the diagrams in Fig.~\ref{fig:rescattering}, we adopt a data-driven approach based on the following parameterization of the hadronic contributions, inspired by the expansion of the correlator of eq.~\eqref{eq:hlambda} as originally done in ref.~\cite{Jager:2012uw}, and worked out in detail in ref.~\cite{Ciuchini:2018anp}:
\begin{eqnarray} 
\label{eq:hv}
H_V^{-} \propto  
 & \frac{m_B^2}{q^2}&  \bigg[ \frac{2m_b}{m_B}\left(C_7^{\rm SM} + h_-^{(0)} \right) \widetilde T_{L -}  
  -  16\pi^2 h_-^{(2)}\, q^4 \bigg] \nnl 
  & + & \left(C_9^{\rm SM} + h_-^{(1)}\right)\widetilde V_{L -}\,, \nnl  
H_V^{+} \propto  
 & \frac{m_B^2}{q^2}&  \bigg[ \frac{2m_b}{m_B}\left(C_7^{\rm SM} + h_-^{(0)} \right) \widetilde T_{L +}  
  -  16\pi^2 \Big(h_{+}^{(0)}  \nnl 
  & + & h_{+}^{(1)}\, q^2 +  h_{+}^{(2)}\, q^4\Big) \bigg] + \left(C_9^{\rm SM} + h_-^{(1)}\right)\widetilde V_{L +}\,, \nnl 
H_V^{0} \propto  
 & \frac{m_B^2}{q^2}&  \bigg[ \frac{2m_b}{m_B}\left(C_7^{\rm SM} + h_-^{(0)} \right) \widetilde T_{L 0}  
  -  16\pi^2 \sqrt{q^2} \Big(h_{0}^{(0)}  \nnl 
  & + & h_{0}^{(1)}\, q^2 \Big) \bigg] + \left(C_9^{\rm SM} + h_-^{(1)}\right)\widetilde V_{L 0}\,.
\end{eqnarray}
This parameterization -- while merely rooted on a phenomenological basis --  has the advantage of making transparent the interplay between hadronic and possible NP contributions. Indeed, the coefficients $h_-^{(0)}$ and $h_-^{(1)}$ have the same effect of a lepton universal shift due to NP in the real part of the Wilson coefficients $C_7$ and $C_9$, respectively. Consequently, the theoretical assumptions on the size of these hadronic parameters crucially affect the extraction of NP contributions to $C_{7,9}$ from global fits. Within the SM, the new measurements in eqs.~\eqref{eq:newBsmmExp}-\eqref{eq:RK22} do not affect the knowledge of the $h_{\lambda}$ coefficients; the most up-to-date data-driven extraction of the hadronic parameters introduced in eq.~\eqref{eq:hv} can be found in Table~1 of ref.~\cite{Ciuchini:2021smi}. See the Appendix for further details regarding the hadronic parameterization employed in the \textit{data driven} approach.

Moving to the analysis of NP, current constraints from direct searches at the LHC reasonably suggest in this context that BSM physics would arise at energies much larger than the electroweak scale. Then, a suitable framework to describe such contributions is given by the SMEFT, in particular by adding to the SM the following dimension-six operators:\footnote{Note that these operators may be generated via renormalization group effects, see, e.g., refs.~\cite{Celis:2017doq,Alasfar:2020mne}.}
\bea \label{eq:SMEFT_op_tree}
O^{LQ^{(1)}}_{2223} & \ \ = \ \ & (\bar{L}_2\gamma_\mu L_2)(\bar{Q}_2\gamma^\mu Q_3)\,, \nonumber \\
O^{LQ^{(3)}}_{2223} & \ \ = \ \ & (\bar{L}_2\gamma_\mu \tau^{A} L_2)(\bar{Q}_2\gamma^\mu\tau^{A} Q_3)\,,\nonumber \\
O^{Qe}_{2322} & \ \ = \ \ & (\bar{Q}_{2}\gamma_\mu Q_{3})(\bar{e}_{2} \gamma^\mu e_{2})\,,\nonumber \\
O^{Ld}_{2223} & \ \ = \ \ & (\bar{L}_2\gamma_\mu L_2)(\bar{d}_2\gamma^\mu d_3)\,,\nonumber \\
O^{ed}_{2223} & \ \ = \ \ & (\bar{e}_2\gamma_\mu e_2)(\bar{d}_2\gamma^\mu d_3)\,,
\eea
where in the above $\tau^{A=1,2,3}$ are Pauli matrices, a sum over $A$ is understood, $L_i$ and $Q_i$ are $SU(2)_{L}$ doublets, $e_i$ and $d_i$ singlets, and flavor indices are defined in the basis where the down-quark Yukawa matrix is diagonal. For concreteness, we normalize SMEFT Wilson coefficients to a NP scale $\Lambda_\mathrm{NP} = 30$ TeV and we only consider NP contributions to muons.\footnote{This choice is mainly motivated by the fact that $B_{s} \to \mu^{+} \mu^{-}$ is one of the key observables of the present study.} 
The matching between the weak effective Hamiltonian and the SMEFT operators implies the following contributions to the SM operators and to the chirality-flipped ones denoted by primes~\cite{Aebischer:2015fzz}: 
\bea \label{eq:SMEFT_matching}
C_{9}^{\rm NP} & \ \ = \ \ & \mathcal{N}_\Lambda\left(C^{LQ^{(1)}}_{22 23} + C^{LQ^{(3)}}_{22 23} + C^{Qe}_{23 22} \right)\,,\nonumber \\
C_{10}^{\rm NP} & \ \ = \ \ & \mathcal{N}_\Lambda\left(C^{Qe}_{23 22} - C^{LQ^{(1)}}_{22 23} - C^{LQ^{(3)}}_{22 23}  \right)\,,\nonumber \\
C^{\prime, \rm{NP}}_{9} & \ \ = \ \ & \mathcal{N}_\Lambda\left(C^{ed}_{22 2 3} + C^{Ld}_{22 2 3} \right)\,,\nonumber \\
C^{\prime, \rm{NP}}_{10} & \ \ = \ \ & \mathcal{N}_\Lambda\left(C^{ed}_{22 2 3} - C^{Ld}_{22 23} \right)\,,
\eea
with $\mathcal{N}_\Lambda = (\pi v^2)/(\alpha_{e}  V_{ts}^{}V_{tb}^* \Lambda_{\rm NP}^2)$. As evident from the above equation, operators $O^{LQ^{(1,3)}}_{2223}$ always enter as a sum. Hence we denote their Wilson coefficient as $C^{LQ}_{2223}$.

\begin{figure*}[!t!]
\ \ \ \
\includegraphics[width=0.45\textwidth]{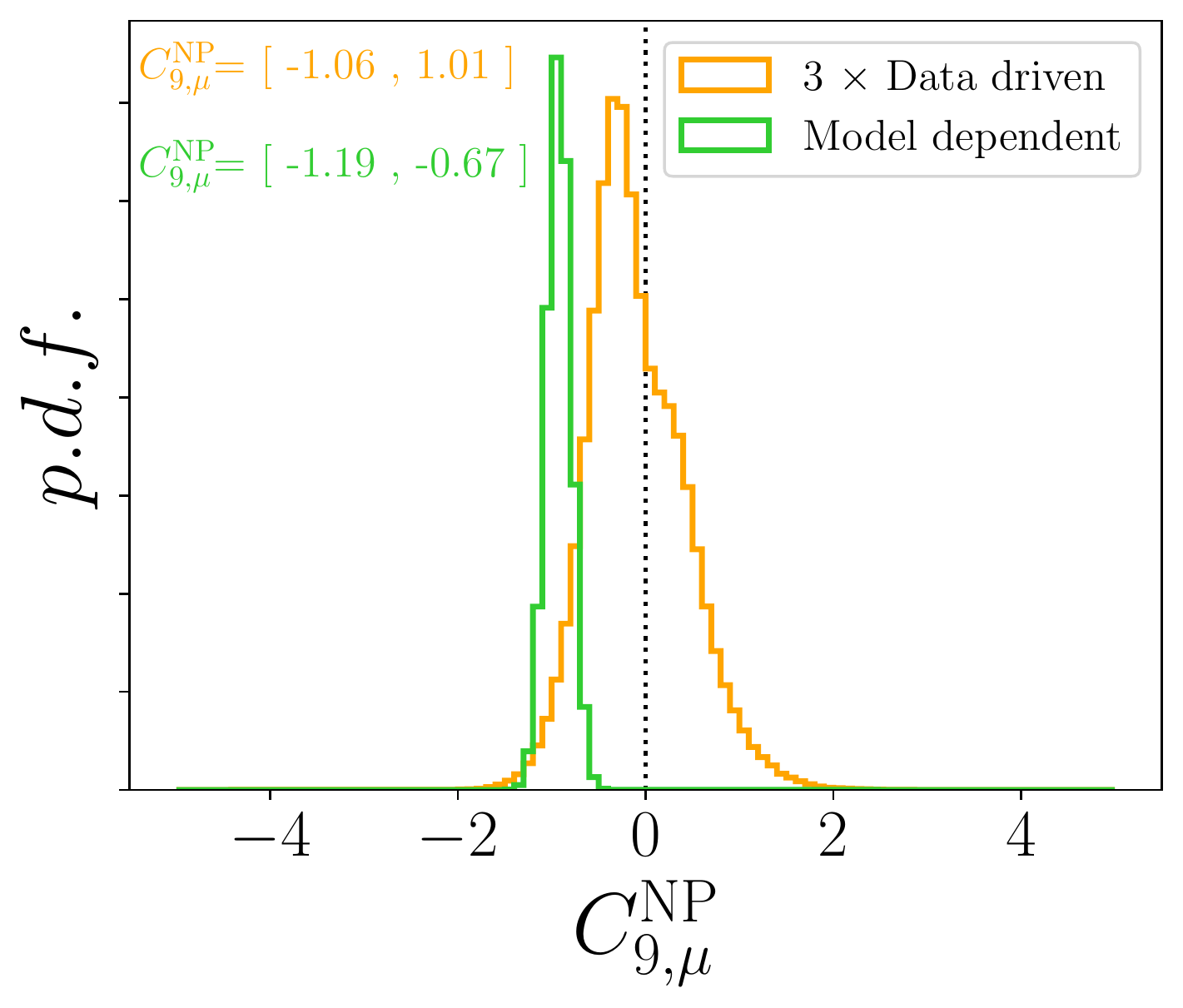}
 \ \ \ \ \ \ 
\includegraphics[width=0.45\textwidth]{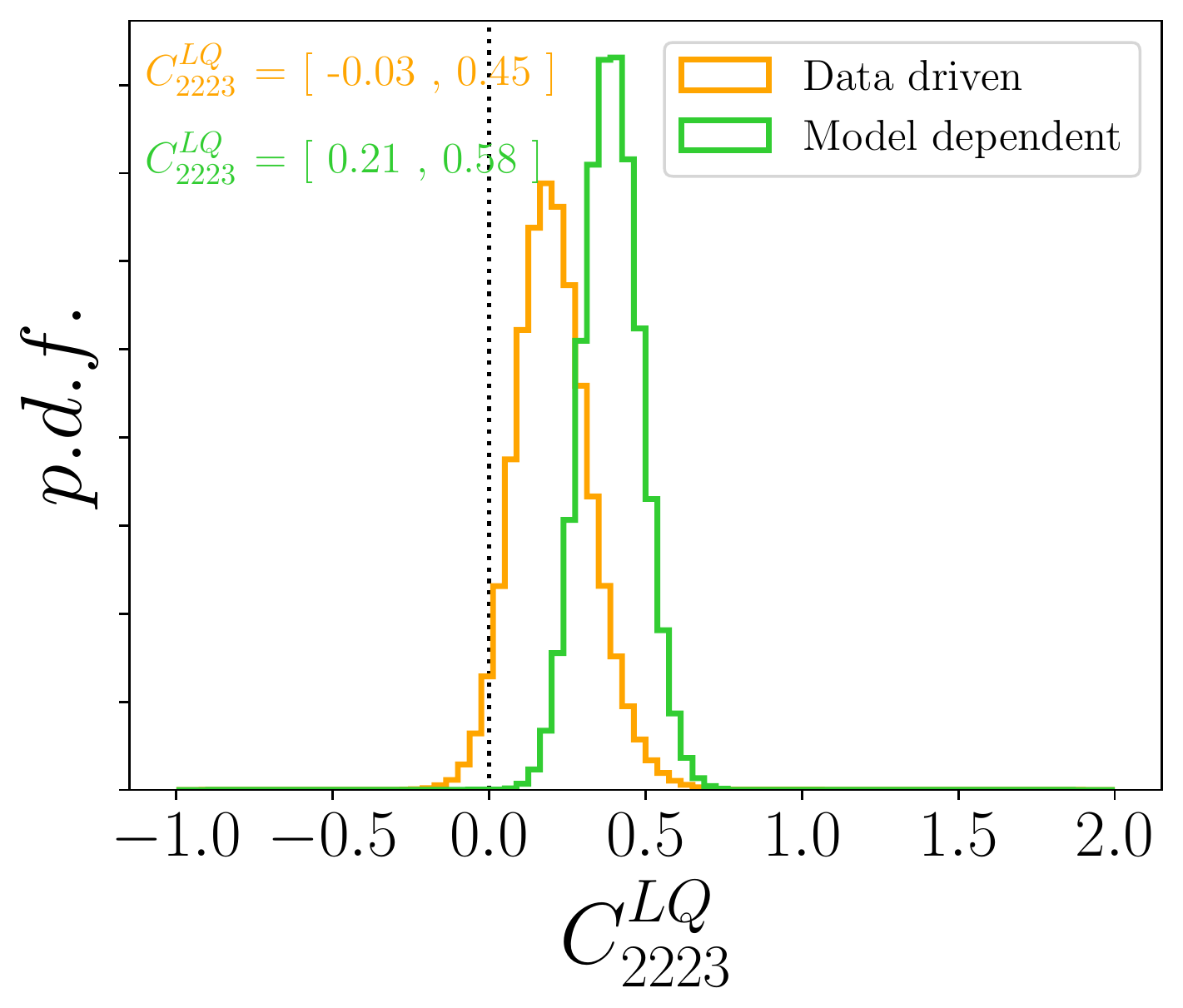}
\caption{\it \underline{Left panel}: Posterior p.d.f. for the NP coefficient $C^{\rm NP}_{9,\mu}$. \underline{Right panel}: Posterior p.d.f. for the SMEFT Wilson coefficient $C^{LQ}_{2223}$. For both panels, we show the p.d.f. in green and orange on the basis of the hadronic approach adopted in the global analysis (see the text for more details).}
\label{fig:C9_CLQ}
\end{figure*}

\begin{figure*}[!t!]
\includegraphics[width=0.46\textwidth]{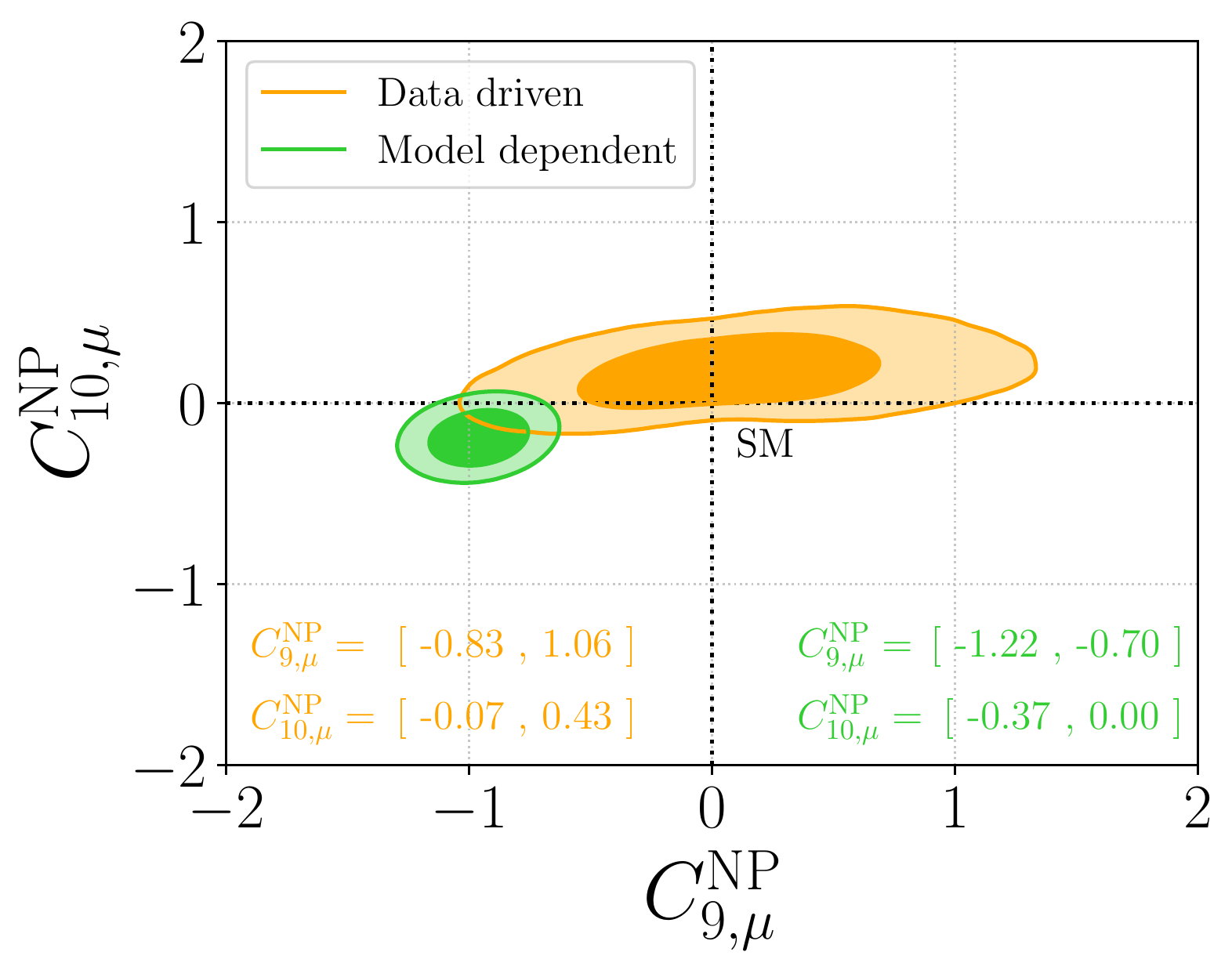}
\ \ 
\includegraphics[width=0.46\textwidth]{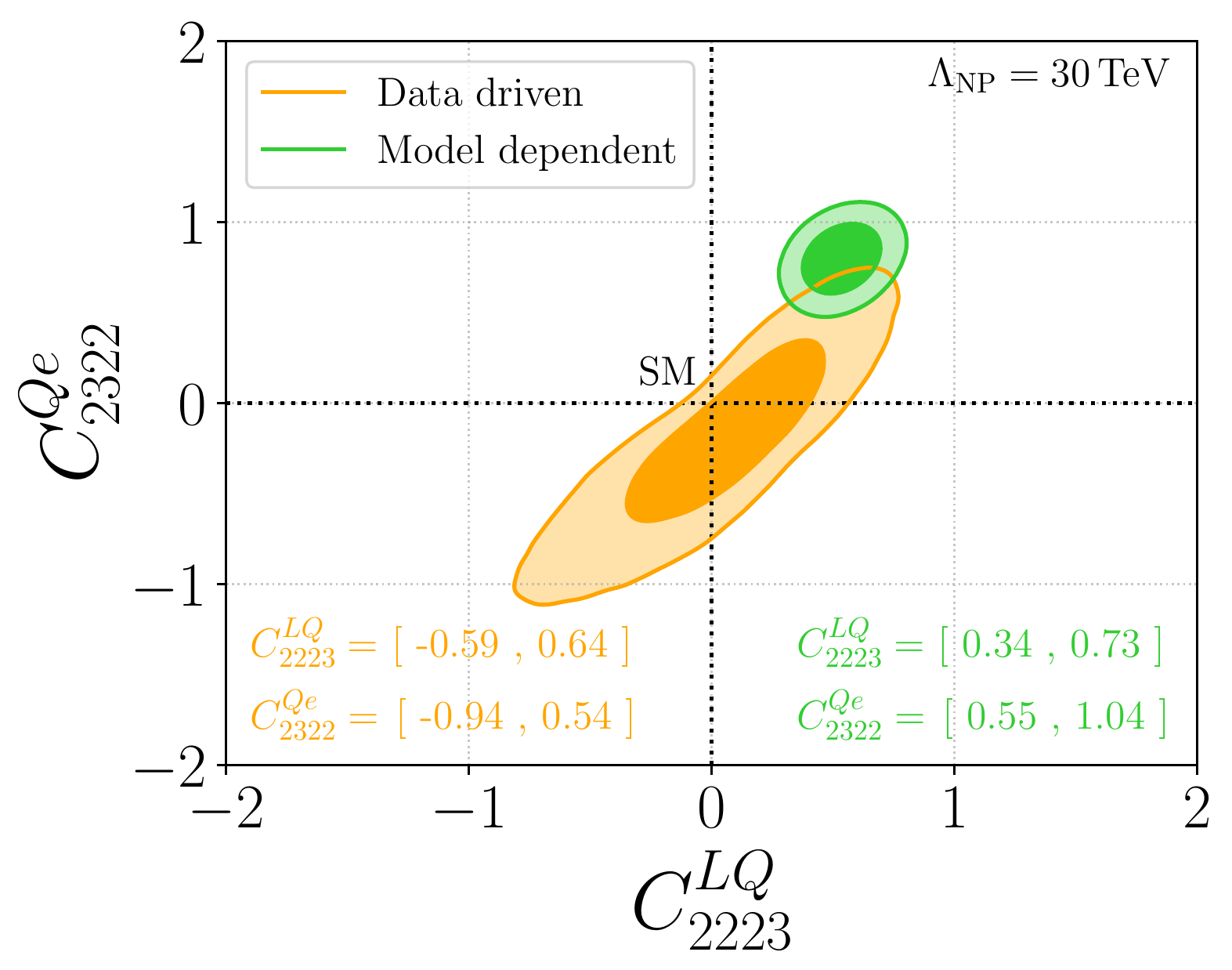}
\caption{\it \underline{Left panel}: Joint posterior p.d.f. for $C^{\rm NP}_{9,\mu}$ and $C^{\rm NP}_{10,\mu}$. \underline{Right panel}: Joint posterior p.d.f. for the SMEFT Wilson coefficients $C^{LQ}_{2223}$ and $C^{Qe}_{2322}$. For both panels, we show $68\%$ and $95\%$ probability regions in green and orange on the basis of the hadronic approach adopted in the global analysis (see the text for more details).}
\label{fig:C9C10_CLQCQe}
\end{figure*}

We perform a Bayesian fit to the data in refs.~\cite{CMS:2014xfa,LHCb:2017rmj,ATLAS:2018cur,CMS:2019bbr,LHCb:2021vsc,LHCb:2020gog,LHCb:2020lmf,LHCb:2015svh,LHCb:2013tgx,LHCb:2015wdu,LHCb:2021zwz,CMS:2020oqb,LHCb:2021xxq,Belle:2016fev,Belle:2019oag,CMS-PAS-BPH-21-006,LHCb:2022qnv,LHCb:2022zyk} employing the \texttt{HEPfit} code \cite{deBlas:2019okz,HEPfit}. For the form factors and input parameters, we follow the same approach used in our previous refs.~\cite{Ciuchini:2015qxb,Ciuchini:2016weo,Ciuchini:2017mik,Ciuchini:2018anp,Ciuchini:2019usw,Ciuchini:2020gvn,Ciuchini:2021smi}. In particular, we use the same inputs as in ref.~\cite{Ciuchini:2021smi}, with the only exception of CKM parameters, which have been updated according to the results of ref.~\cite{UTfit:2022hsi}. We compute $B \to K^{(*)}\ell^+ \ell^-$ and $B_s \to \phi \ell^+ \ell^-$ decays using QCD factorization \cite{Beneke:2004dp}. 
    
As already mentioned discussing Fig.~\ref{fig:rescattering}, a global analysis of $b \to s \ell^{+} \ell^{-}$ transitions can be  sensitive to hadronic contributions that are difficult to compute from first principles and that can yield important phenomenological effects. Therefore, in what we denote below as \textit{data driven} scenario, we assume a flat prior in a sufficiently large range for the $h_\pm^{(0,1,2)}$ and $h_0^{(0,1)}$ parameters, which are then determined from data simultaneously with the NP coefficients.\footnote{As in ref.~\cite{Ciuchini:2021smi}, we assume exact SU(3) flavor for the $h$ parameters and add additional ones for $B\to K$.} To clarify the phenomenological relevance of charming penguins, we compare the results of the \textit{data driven} approach against what we denote instead as \textit{model dependent} treatment of hadronic uncertainties, in which we assume that the contributions generated by the diagrams in Fig.~\ref{fig:rescattering}~\textbf{(b)} (or \textbf{(c)}) are negligible and that the correlator in eq.~\eqref{eq:hlambda} is well described by the approach of refs.~\cite{Khodjamirian:2010vf,Khodjamirian:2012rm,Bobeth:2017vxj,Chrzaszcz:2018yza,Gubernari:2020eft,Gubernari:2022hxn}, yielding a subleading effect to the hadronic effects computable in QCD factorization. See the Appendix for further details regarding the parameterization of hadronic contributions employed in the \textit{model dependent} approach.
    
In both approaches to QCD long-distance effects, we obtain a sample of the posterior joint probability density function (p.d.f.) of SM parameters, including form factors, and, in the \textit{data driven} scenario, $h_{\lambda}$ parameters, together with NP Wilson coefficients. From each posterior p.d.f. we compute the highest probability density intervals (HPDIs), which represent our best knowledge of the model parameters after the new measurements. We also perform model comparison using the information criterion~\cite{IC}, defined as:
    \begin{equation}\label{eq:IC}
   IC \equiv -2 \overline{\log \mathcal{L}} \, + \, 4 \sigma^{2}_{\log \mathcal{L}} \,,
\end{equation}
where the first and second terms are the mean and variance of the log-likelihood posterior distribution. The first term measures the quality of the fit, while the second one is related to the effective degrees of freedom involved, penalizing more complicated models. Models with smaller IC should then be preferred~\cite{BayesFactors}. While the posterior distributions for SM parameters are unaffected by LUV measurements, the SM IC of course depends on the latter: indeed, the SM in the \textit{data driven} scenario provides an excellent description of current data, leading to slightly negative values of $\Delta\mathrm{IC} \equiv \mathrm{IC}_\mathrm{SM}-\mathrm{IC}_\mathrm{NP}$. Conversely, the agreement of the SM with angular observables remains poor in the \textit{model dependent} approach, implying for this case large values of $\Delta\mathrm{IC}$, signaling a statistically significant preference for NP. 

\begin{table}[!t]
\centering
\renewcommand{\arraystretch}{1.5}
{\footnotesize
\begin{tabular}{|c|cc|}
\hline
& 95\% HPDI & $\Delta IC$ \\
\hline
\multirow{2}{*}{$ C_{9,\mu}^{\rm NP} $}
& [ -1.06 , 1.01 ] & -2.4 \\
&\cellcolor{Gray} [ -1.19 , -0.67 ]  &\cellcolor{Gray} 43  \\
\hline
\multirow{2}{*}{$ \{C_{9,\mu}^{\rm NP},C_{10,\mu}^{\rm{NP}}\} $}
& \{[ -0.83 , 1.06 ], [ -0.07 , 0.43 ]\}  & -3.4 \\
&\cellcolor{Gray} \{[ -1.22 , -0.70 ], [ -0.37 , 0.00 ]\}  &\cellcolor{Gray}  41 \\
\hline
\multirow{2}{*}{$ \{C_{9,\mu}^{\rm NP},C_{9,\mu}^{\prime, \rm{NP}}\} $}
& $\{[ -1.06 , 1.40 ], [ -2.20 , 1.31 ],$ & -4.1 \\
&\cellcolor{Gray} $\{[ -1.33 , -0.79 ], [ 0.08 , 0.88 ],$  &\cellcolor{Gray}  45 \\
\hline
\multirow{2}{*}{$ \{C_{9,\mu}^{\rm NP},C_{10,\mu}^{\prime, \rm{NP}}\} $}
& $\{[ -1.07 , 1.20 ], [ -0.28 , 0.20 ],$ & -5.1 \\
&\cellcolor{Gray} $\{[ -1.34 , -0.77 ], [ -0.39 , 0.02 ],$  &\cellcolor{Gray} 41 \\
\hline
& $\{[ -0.90 , 1.49 ], [ -0.15 , 0.62 ],$ & \multirow{2}{*}{-8.1} \\ 
$\{C_{9,\mu}^{\rm NP},C_{10,\mu}^{\rm{NP}},$
& $[ -2.27 , 1.18 ], [ -0.33 , 0.47 ]\}$&\\
$C_{9,\mu}^{\prime, \rm NP},C_{10,\mu}^{\prime, \rm{NP}}\}$ 
& \cg $\{[ -1.38 , -0.82 ], [ -0.39 , 0.02 ],$ & \cg \\
& \cg $[ -0.49 , 0.79 ], [ -0.46 , 0.17 ]\}$& \cg\multirow{-2}{*}{57}  \\
\hline
\end{tabular}
}
\caption{\em HPDI for the Wilson coefficients of the low-energy weak Hamiltonian in all the considered NP scenarios along with the corresponding $\Delta IC$. White rows correspond to results obtained in the \textit{data driven} scenario, while \textit{model dependent} scenario results are shaded in gray. See the text for the definition of the two scenarios. 
\label{tab:WC_WET}}
\end{table}

\begin{table}[!t]
\centering
\renewcommand{\arraystretch}{1.5}
{\footnotesize
\begin{tabular}{|c|cc|}
\hline
&  95\% HPDI & $\Delta IC$ \\
\hline
\multirow{2}{*}{$ C^{LQ}_{2223} $}
& [ -0.03 , 0.45 ] & -1.6 \\
&\cellcolor{Gray} [ 0.21
 , 0.58 ] & \cellcolor{Gray} 3 \\
\hline
\multirow{2}{*}{$ \{C^{LQ}_{2223},C^{Qe}_{2322}\} $}
& \{[ -0.59 , 0.64 ], [ -0.94 , 0.54 ]\} & -3.4 \\
&\cellcolor{Gray} \{[ 0.34 , 0.73 ] , [ 0.55 , 1.04 ]\} &\cellcolor{Gray} 41 \\
\hline
\multirow{2}{*}{$ \{C^{LQ}_{2223},C^{ed}_{2223}\} $}
& \{[ -0.03 , 0.48 ], [ -0.39 , 0.32 ]\} & -4.0 \\
&\cellcolor{Gray} \{[ 0.24 , 0.63 ], [ -0.95 , -0.09 ]\} &\cellcolor{Gray} 6 \\
\hline
\multirow{2}{*}{$ \{C^{LQ}_{2223},C^{Ld}_{2223}\} $}
& \{[ -0.06 , 0.65 ], [ -0.24 , 0.49 ]\} &  -5.1 \\
&\cg \{[ 0.18 , 0.57 ], [ -0.14 , 0.23 ]\} &\cg -2 \\
\hline
& $\{[ -0.88 , 0.78 ], [ -1.26 , 0.57 ],$ & \multirow{2}{*}{-8.1} \\ 
$\{C^{LQ}_{2223},C^{Qe}_{2322},$
& $[ -0.76 , 1.58 ],  [-0.98 , 1.64 ]\}$&\\
$C^{Ld}_{2223},C^{ed}_{2223}\}$ 
& \cg $\{[ 0.43 , 0.84 ], [ 0.62 , 1.16 ],$ & \cg \\
& \cg $[ -0.50 , 0.10 ], [ -0.64 , 0.63 ]\}$& \cg\multirow{-2}{*}{57}  \\
\hline
\end{tabular}
}
\caption{\em Same as Tab.~\ref{tab:WC_WET} for SMEFT Wilson coefficients. 
\label{tab:WC_SMEFT}}
\end{table}

We now discuss several NP configurations, in order of increasing complexity. We start by allowing a single nonvanishing NP Wilson coefficient, either $C_{9,\mu}^\mathrm{NP}$, defined in the low-energy weak Hamiltonian, or the Wilson coefficient $C_{2223}^{LQ}$, belonging to the SMEFT. The p.d.f.s for the two NP Wilson coefficients are reported in Fig.~\ref{fig:C9_CLQ}, while the corresponding numerical results for the $95\%$ HPDIs are reported in the first row of Tables \ref{tab:WC_WET} and \ref{tab:WC_SMEFT}. As anticipated above, no significant preference for NP is seen in the \textit{data driven} scenario, while NP contributions are definitely needed in the \textit{model dependent} scenario, with a clear preference for $C_{9,\mu}^\mathrm{NP}\neq 0$. 

Figure \ref{fig:C9C10_CLQCQe} displays the allowed regions in the $C_{9,\mu}^\mathrm{NP}-C_{10,\mu}^\mathrm{NP}$ and $C_{2223}^{LQ}-C_{2322}^{Qe}$ planes, while the corresponding HPDIs are reported in the second row of Tables \ref{tab:WC_WET} and \ref{tab:WC_SMEFT} respectively. Again, no evidence for NP is seen in the \textit{data driven} case, while clear evidence for a nonvanishing $C_{9,\mu}^\mathrm{NP}$ appears in the \textit{model dependent} approach. Deviations from zero of $C_{10,\mu}^\mathrm{NP}$ are strongly constrained by $\mathrm{BR}(B_s \to \mu^+ \mu^-)$, corresponding to the strong correlation $C_{2223}^{LQ} \sim C_{2322}^{Qe}$ seen in the right panel of Fig.~\ref{fig:C9C10_CLQCQe}.

\begin{figure}[t!]
    \centering
    \includegraphics[width=0.45\textwidth]{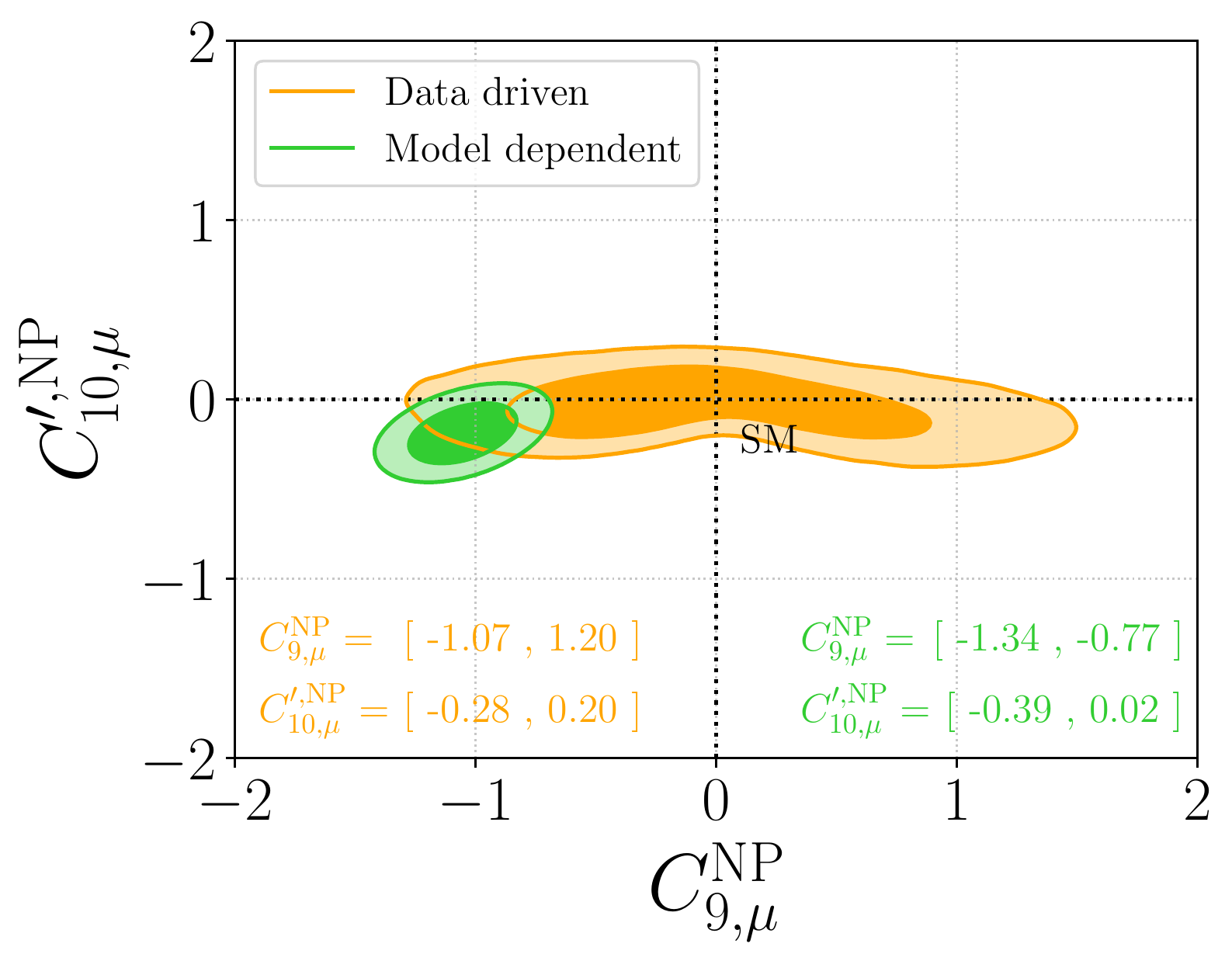}
    \caption{\em Joint posterior p.d.f. for $C^{\rm NP}_{9,\mu}$ and $C^{\prime \rm NP}_{10,\mu}$. We show $68\%$ and $95\%$ probability regions in green and orange on the basis of the hadronic approach adopted in the global analysis (see the text for more details).}
    \label{fig:C9C10p}
\end{figure}

Next, we consider NP models in which right-handed $b\to s$ transitions arise. In the weak effective Hamiltonian, we allow for nonvanishing $C_{9,\mu}^\mathrm{NP}$ and $C_{9,\mu}^{\prime,\mathrm{NP}}$ or $C_{10,\mu}^{\prime,\mathrm{NP}}$. In particular, in Fig.\ref{fig:C9C10p} we present the results of the fit in the $C_{9,\mu}^\mathrm{NP}-C_{10,\mu}^{\prime,\mathrm{NP}}$ case, which we considered in ref.~\cite{Ciuchini:2019usw} as the best fit one in view of the deviation from one of the ratio $R_K/R_{K^*}$~\cite{Hiller:2014yaa}. With the current experimental situation, this is not the case anymore, and $C_{10,\mu}^{\prime,\mathrm{NP}}$ is again strongly constrained by $\mathrm{BR}(B_s \to \mu^+ \mu^-)$. In the SMEFT, we consider nonvanishing $C^{LQ}_{2223}$ and $C^{ed}_{2223}$  or $C^{Ld}_{2223}$. The numerical results for the NP coefficients can be found in the third and fourth rows of Tables \ref{tab:WC_WET} and \ref{tab:WC_SMEFT}.  

\begin{figure*}[t]
  \centering
\includegraphics[width=0.95\textwidth]{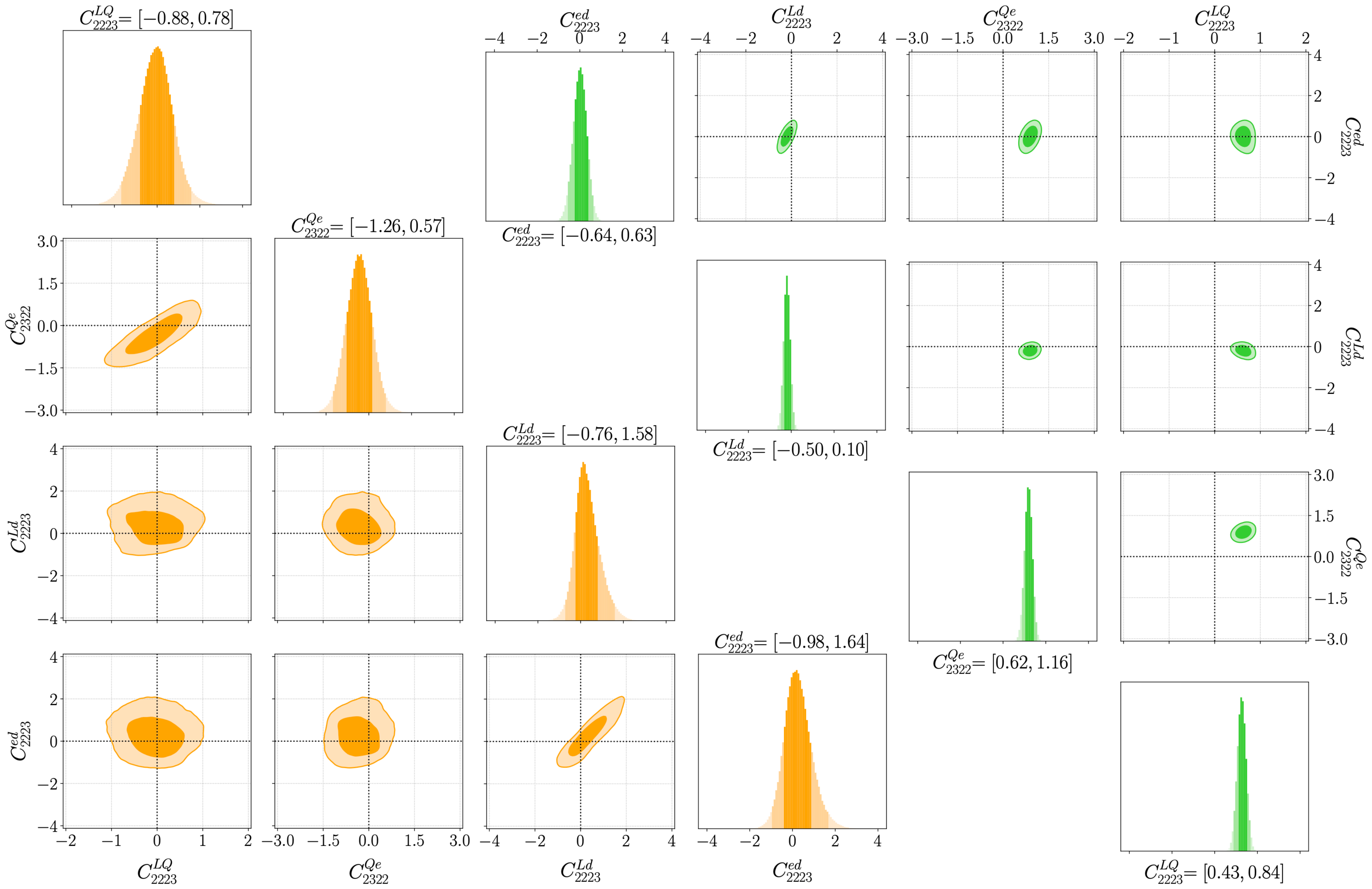}
  \caption{\it Two- and one-dimensional marginalized joint p.d.f. for the set of SMEFT Wilson coefficients $C^{LQ}_{2223}$, $C^{Qe}_{2322}$,  $C^{ed}_{2223}$ and $C^{Ld}_{2223}$. For both panels, we show the $68\%$ and $95\%$ probability regions in green and orange on the basis of the hadronic approach adopted in the global analysis (see the text for more details).}
  \label{fig:Call}
\end{figure*}

\begin{figure*}[t!]
  \centering
\includegraphics[width=0.39\textwidth]{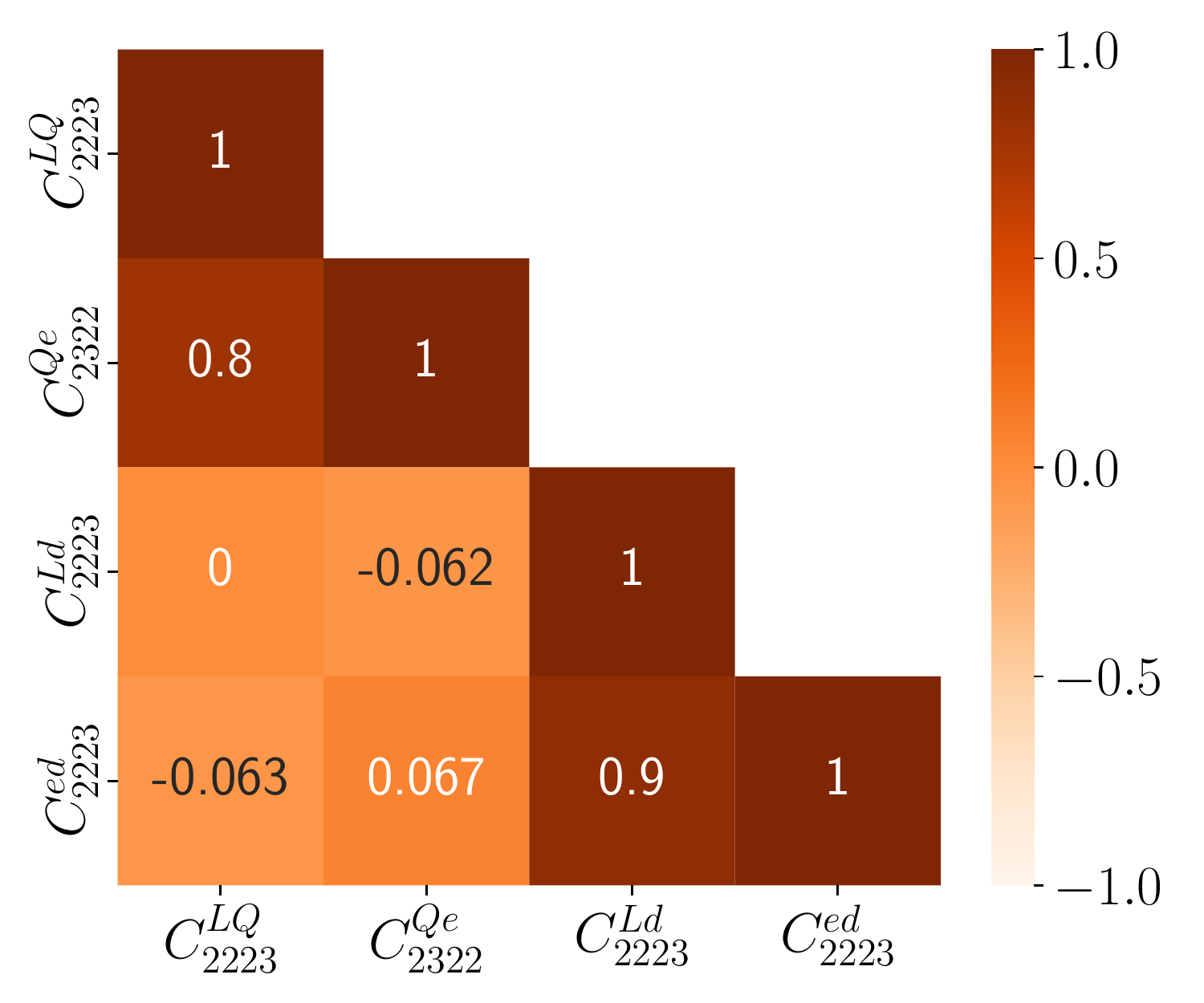}
\includegraphics[width=0.39\textwidth]{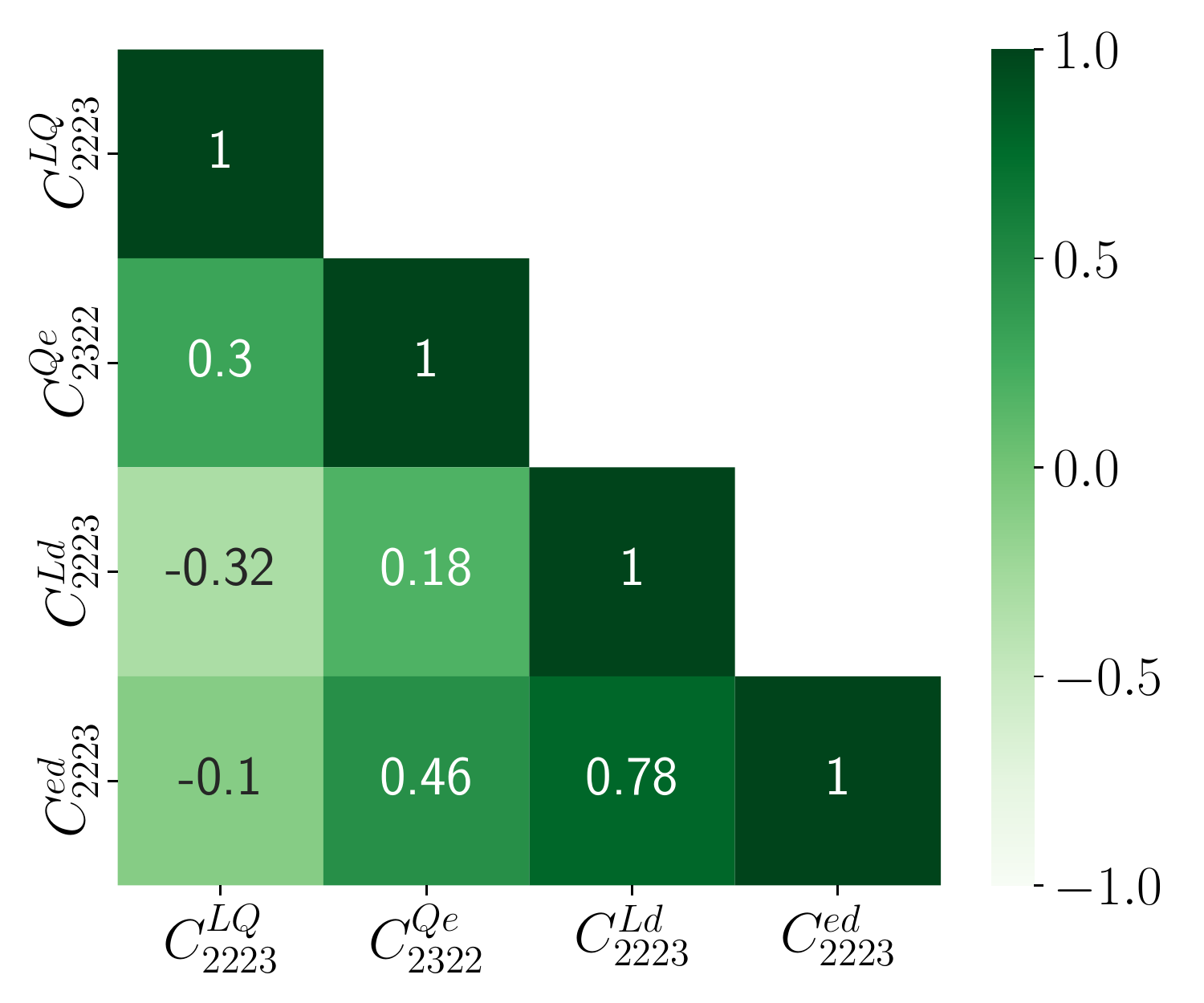}
  \caption{\it Correlation matrix of the Wilson coefficients of the SMEFT operators studied in this work under the ``data driven'' (left panel, orange) and the ``model dependent'' (right panel, green)  approaches to hadronic uncertainties in our global analysis.}
  \label{fig:Call_corr}
\end{figure*}

Finally, we present the results of a combined fit in which all the four NP Wilson coefficients considered above are allowed to float simultaneously, namely $C_{2223}^{LQ}$, $C_{2322}^{Qe}$, $C_{2223}^{Ld}$ and  $C_{2223}^{ed}$, or equivalently, in the language of the weak effective Hamiltonian, $C_{9,\mu}^\mathrm{NP}$, $C_{10,\mu}^\mathrm{NP}$ and the corresponding operators with right-handed quark currents $C_{9,\mu}^{\prime,\mathrm{NP}}$, $C_{10,\mu}^{\prime,\mathrm{NP}}$. Several interesting features emerge in this fit. First, the updated experimental value of $\mathrm{BR}(B_s \to \mu^+ \mu^-)$ forces $C_{10,\mu}^\mathrm{NP}$ and $C_{10,\mu}^{\prime,\mathrm{NP}}$ to be small, corresponding to the correlations visible in the two-dimensional projections on the $C_{2223}^{LQ}$ vs $C_{2322}^{Qe}$ and $C_{2223}^{Ld}$ vs $C_{2223}^{ed}$ planes and reported in Fig.~\ref{fig:Call_corr}. Second, the SM point is well inside the $68\%$ probability regions in the \textit{data driven} approach, while in the \textit{model dependent} scenario there is evidence of a nonvanishing $C_{9,\mu}^\mathrm{NP}$, or equivalently of a nonvanishing $C_{2223}^{LQ} \sim C_{2322}^{Qe}$, stemming from BRs and angular distributions of $b \to s \mu^+ \mu^-$ transitions. In the \textit{data driven} scenario the latter are reproduced thanks to the charming penguin contributions. Eventually, notice that the allowed ranges for NP coefficients are much larger in the \textit{data driven} scenario since the uncertainties on charming penguins leak into the determination of NP Wilson coefficients. 

\begin{figure}[t!]
    \centering
    \includegraphics[width=0.4\textwidth]{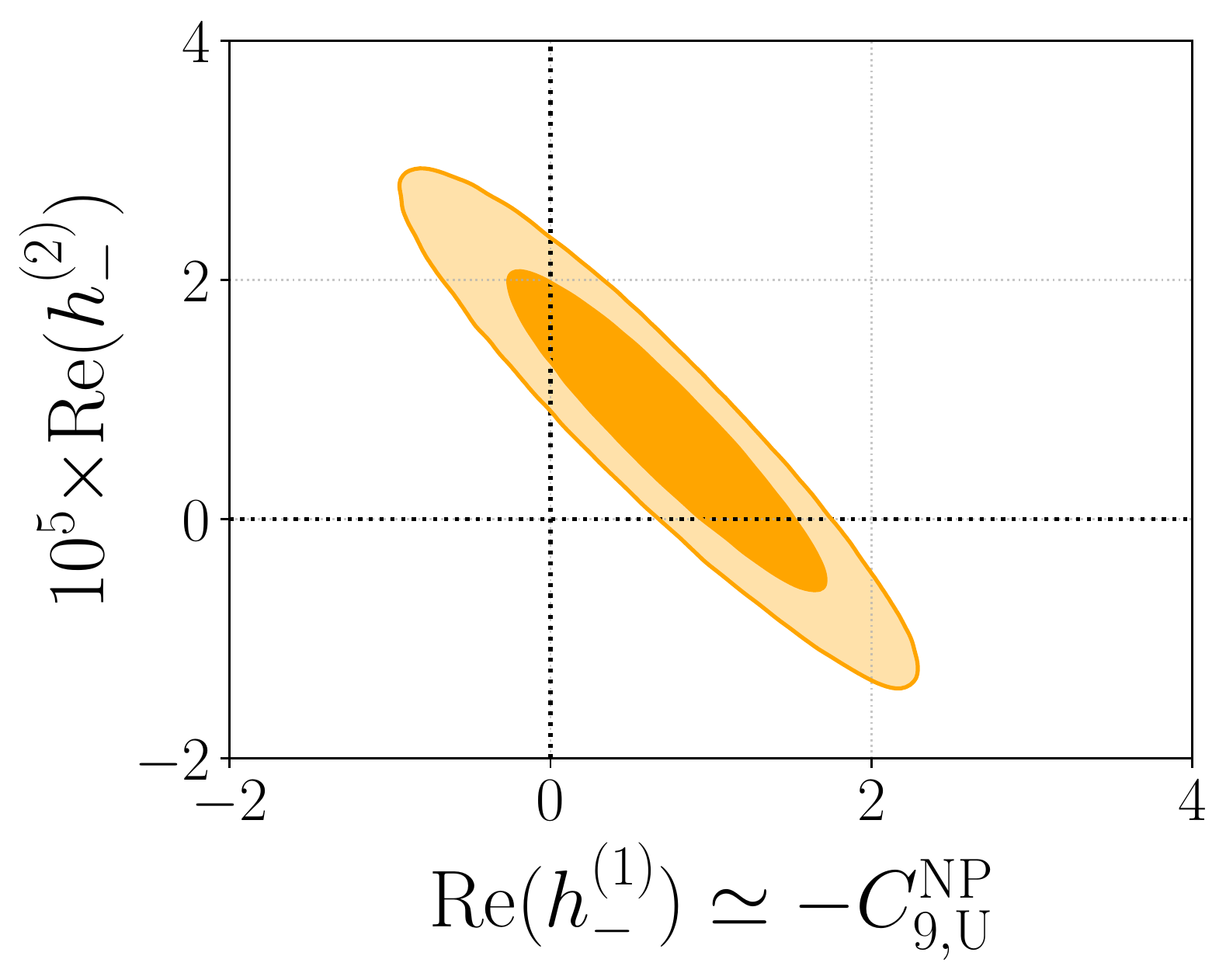}
    \caption{\em Joint posterior p.d.f. for $\mathrm{Re}(h_-^{(1)})$ and $\mathrm{Re}(h_-^{(2)})$ in a SM fit in the ``data driven'' scenario. Darker (lighter) regions correspond to $68\%$ $(95\%)$ probability. Notice that according to our hadronic parameterization given in eq.~\eqref{eq:hv}, $\mathrm{Re}(h_-^{(1)})$ can be reinterpreted as a lepton universal NP contribution, $C_{\rm 9,U}^{\rm NP}$.}
    \label{fig:hminus}
\end{figure}

Before concluding, we comment briefly on the possibility of a lepton universal NP contribution to $C_9$, that we denote here $C^{\rm NP}_{\rm 9,U}$, affecting only absolute BRs and angular distributions of $b \to s \ell^+ \ell^-$ decays, but leaving LUV ratios as in the SM. This possibility was already discussed in detail in ref.~\cite{Ciuchini:2021smi}, and the experimental situation has not changed since then. Therefore, we just summarize here the main findings of ref.~\cite{Ciuchini:2021smi} for the reader's convenience. Performing a fit to experimental data within the SM in the \textit{data driven} scenario, one finds that several $h_\lambda$ parameters are determined to be different from zero at $95\%$ probability, supporting the picture of sizable rescattering in charming penguin amplitudes (see Table 1 in ref.~\cite{Ciuchini:2021smi}). In particular, there is an interesting correlation between $\mathrm{Re}(h_-^{(1)})\simeq - C^{\rm NP}_{\rm 9,U}$ and $\mathrm{Re}(h_-^{(2)})$, as is evident from Fig.~\ref{fig:hminus}.~\footnote{To identify $\mathrm{Re}(h_-^{(1)})$ as $C^{\rm NP}_{\rm 9,U}$, we work in the flavour $SU(3)_{F}$ symmetric limit, in which the same hadronic contribution affects both $B\to K^*$ and $B_s\to\phi$ transitions (see the Appendix for further details); moreover, for the sake of simplicity, we focus only on these two channels and do not take in consideration additional correlations with other hadronic parameters that similarly mimic the effect of $C^{\rm NP}_{\rm 9,U}$ in $B\to K$ transitions.} Data definitely require a nonvanishing combination of the two parameters; if charming penguins are treated à la \cite{Khodjamirian:2010vf,Khodjamirian:2012rm,Bobeth:2017vxj,Chrzaszcz:2018yza,Gubernari:2020eft,Gubernari:2022hxn}, $\mathrm{Re}(h_-^{(2)})$ is put to zero and $\mathrm{Re}(h_-^{(1)})$ is identified with a lepton universal contribution $C^{\rm NP}_{\rm 9,U}$, leading to an evidence of NP inextricably linked to the assumptions on charming-penguin amplitudes.

Summarizing, we performed a Bayesian analysis of possible LUV NP contributions to $b\to s \ell^+ \ell^-$ transitions in view of the very recent updates on BR$(B_{(d,s)} \to \mu^+ \mu^-)$ by the CMS collaboration \cite{CMS-PAS-BPH-21-006} and on $R_K$ and $R_{K^*}$ by the LHCb collaboration \cite{LHCb:2022qnv,LHCb:2022zyk}. As pointed out in refs.~\cite{Jager:2012uw,Jager:2014rwa,Ciuchini:2015qxb,Ciuchini:2016weo,Ciuchini:2017mik,Ciuchini:2018anp,Ciuchini:2019usw,Ciuchini:2020gvn,Ciuchini:2021smi}, the NP sensitivity of these transitions is spoilt by possible long-distance effects, see~Fig.~\ref{fig:rescattering}. 
Thus, in the \textit{data driven} scenario we determined simultaneously hadronic contributions, parameterized according to eq.~\eqref{eq:hlambda}, and NP Wilson coefficients, finding no evidence for LUV NP. Conversely, evidence for NP contributions is found if charming penguins are assumed to be well described by the approach of refs.~\cite{Khodjamirian:2010vf,Khodjamirian:2012rm,Bobeth:2017vxj,Chrzaszcz:2018yza,Gubernari:2020eft,Gubernari:2022hxn}, as reported in Tables \ref{tab:WC_WET} and \ref{tab:WC_SMEFT}. 

Finally, we considered the case of a lepton universal NP contribution to $C_9$, which is phenomenologically equivalent to the effect of $h_-^{(1)}$ in our \textit{data driven} analysis, confirming our previous findings in ref.~\cite{Ciuchini:2021smi}: in the context of the \textit{data driven} approach, we found several hints of nonvanishing $h_\lambda^i$ parameters, but no evidence of a nonvanishing Re$(h_-^{(1)})\simeq - C^{\rm NP}_{\rm 9,U}$; evidence for $C^{\rm NP}_{\rm 9,U}$ only arises in the \textit{model dependent} scenario in which all genuine hadronic contributions are phenomenologically negligible. Future improvements in theoretical calculations and in experimental data will hopefully allow clarifying this last point.

\section*{Appendix}

In this Appendix, we give further details regarding the parameterizations employed for the hadronic contributions in the \textit{data driven} and \textit{model dependent} approaches in each of the two main decays investigated in this work, namely $B \to K^* \ell \ell$ and $B \to K \ell \ell$, and how these approaches are related to each other. Concerning the third process discussed in this work, namely $B_s \to \phi \ell \ell$, we work under the assumption of $SU(3)_{F}$ symmetry, i.e., we consider the same hadronic contributions to $B \to K^* \ell \ell$ and $B_s \to \phi \ell \ell$. This choice is justified by the fact that it is not possible with current data to single out any $SU(3)_{F}$-breaking effect from $B_s\to\phi \ell\ell$, see our previous work in ref.~\cite{Ciuchini:2021smi} for a detailed analysis on this matter. Starting from the \textit{model dependent} approach in the $B \to K^*$ mode, we follow the definition of ref.~\cite{Khodjamirian:2010vf} and give the hadronic contributions as helicity-dependent shifts in $C_{9,i}$:
\begin{equation}\label{eq:DC9_LCSR}
\Delta C_{9,i} (q^2) = \frac{r_{1,i}\left(1-\frac{\bar q^2}{q^2}\right) + \Delta C_{9,i} (\bar q^2) \frac{\bar q^2}{q^2}}{1+r_{2,i}\frac{\bar q^2-q^2}{m^2_{J/\psi}}}\,.
\end{equation}

\begin{table}[!t]
\centering
\renewcommand{\arraystretch}{1.5}
{\footnotesize
\begin{tabular}{|c|c|c|c|}
\hline
\phantom{aa}$\vert\Delta C_{9,1}(q^2)\vert$\phantom{aa} &  \phantom{aa}\emph{data driven}\phantom{aa} & \emph{model dependent} & QCDF \\
\hline
$\vert\Delta C_{9,1}(1.0)\vert$ &   [1.58,   5.43] & [0.88,   1.08] & [0.45, 0.55] \\
\hline
$\vert\Delta C_{9,1}(1.5)\vert$ &   [1.23,   4.08] & [0.59,   0.73] & [0.38, 0.50] \\
\hline
$\vert\Delta C_{9,1}(2.0)\vert$ &   [1.00,   3.32] & [0.45,   0.56] & [0.35, 0.45] \\
\hline
$\vert\Delta C_{9,1}(2.5)\vert$ &   [0.82,   2.75] & [0.37,   0.46] & [0.32, 0.43] \\
\hline
$\vert\Delta C_{9,1}(3.0)\vert$ &   [0.66,   2.28] & [0.31,   0.40] & [0.32, 0.42] \\
\hline
$\vert\Delta C_{9,1}(3.5)\vert$ &   [0.53,   1.88] & [0.28,   0.36] & [0.31, 0.40] \\
\hline
$\vert\Delta C_{9,1}(4.0)\vert$ &   [0.44,   1.58] & [0.26,   0.34] & [0.31, 0.42] \\
\hline
$\vert\Delta C_{9,1}(4.5)\vert$ &   [0.41,   1.41] & [0.25,   0.33]& [0.31, 0.43] \\
\hline
$\vert\Delta C_{9,1}(5.0)\vert$ &   [0.45,   1.39] & [0.25,   0.33] & [0.32, 0.43] \\
\hline
$\vert\Delta C_{9,1}(5.5)\vert$ &   [0.57,   1.52] & [0.25,   0.33] & [0.32, 0.45] \\
\hline
$\vert\Delta C_{9,1}(6.0)\vert$ &   [0.75,   1.75] & [0.26,   0.35] & [0.33, 0.46] \\
\hline
$\vert\Delta C_{9,1}(6.5)\vert$ &   [0.95,   2.03] & [0.26,   0.35] & [0.33, 0.48] \\
\hline
$\vert\Delta C_{9,1}(7.0)\vert$ &   [1.15,   2.36] & [0.34,   0.43] & [0.36, 0.50] \\
\hline
$\vert\Delta C_{9,1}(7.5)\vert$ &   [1.36,   2.70] & [0.55,   0.66] & [0.40, 0.55] \\
\hline
$\vert\Delta C_{9,1}(8.0)\vert$ &   [1.55,   3.06] & [0.86,   1.01] & [0.47, 0.60] \\
\hline
\end{tabular}
}
\caption{\em 68\% HPDI for the hadronic contribution $\vert\Delta C_{9,1}(q^2)\vert$ entering in $B \to K^*$ and $B \to \phi$ transitions at different values of $q^2$, both in the \emph{data driven} and the \emph{model dependent} approaches. In the last column, we also report the expected size of the contributions coming from QCDF.
\label{tab:DC91_comparison}}
\end{table}
In our fits, all the involved parameters are considered real according to the way they have been defined and computed in ref.~\cite{Khodjamirian:2010vf}, namely by performing a Wick rotation to the Euclidean space in order to compute the light cone sum rule. In particular, they are considered flatly distributed according to the ranges given in Table 2 of the same reference, for $\bar q^2=1$. As discussed in ref.~\cite{Ciuchini:2015qxb}, the relation between this parameterization and the one employed for the \textit{data driven} approach is given by:
\begin{eqnarray}
\Delta C_{9,1}(q^2) & = & - \frac{16 m_{B}^3(m_B+m_{K^{*}})\pi^2}{\sqrt{\lambda(q^2)}V(q^2) q^2} (h_{-}(q^2)-h_{+}(q^2))\nonumber\\
\Delta C_{9,2}(q^2)  & = &  - \frac{16 m_{B}^3\pi^2}{(m_B+m_{K^{*}})A_{1}(q^2)q^2} (h_{-}(q^2)+h_{+}(q^2)) \nonumber \\
\Delta C_{9,3}(q^2) & = & 
\frac{64 \pi^2 m_B^3 m_{K^{*}} \sqrt{q^2} (m_B + m_{K^*})}{\lambda(q^2) A_{2}(q^2) q^2} \, h_{0}(q^2) \nonumber \\ 
& \ - \ &\frac{16 m_{B}^3(m_B+m_{K^{*}})(m_B^2-q^2-m_{K^{*}}^2)\pi^2}{\lambda(q^2) A_{2}(q^2) q^2} \nonumber \\  
 & \ \times \ & (h_{-}(q^2)+h_{+}(q^2)) \,,
\label{eq:gtildes}
\end{eqnarray}
where we have introduced the helicity functions $h_\lambda(q^2)$. These functions have been defined in such a way that, in the helicity amplitudes shown in Eq.~\eqref{eq:hv}, the coefficients $h_-^{(0)}$ and $h_-^{(1)}$ have the same effect of a NP lepton universal shift in the real part of $C_7$ and $C_9$, namely
\bea\label{eq:hla_BtoKst}
h_-(q^2) &=& -\frac{m_b}{8\pi^2 m_B} \widetilde T_{L -}(q^2) h_-^{(0)} -\frac{\widetilde V_{L -}(q^2)}{16\pi^2 m_B^2} h_-^{(1)} q^2 \nonumber\\
&&+ h_-^{(2)} q^4+{\cal O}(q^6)\,,\nnl
h_+(q^2) &=&  -\frac{m_b}{8\pi^2 m_B} \widetilde T_{L +}(q^2) h_-^{(0)} -\frac{\widetilde V_{L +}(q^2)}{16\pi^2 m_B^2}  h_-^{(1)} q^2 \nonumber\\
&&+ h_+^{(0)} + h_+^{(1)}q^2 + h_+^{(2)} q^4+{\cal O}(q^6)\,, \nnl
h_0(q^2) &=& -\frac{m_b}{8\pi^2 m_B} \widetilde T_{L 0}(q^2) h_-^{(0)} -\frac{\widetilde V_{L 0}(q^2)}{16\pi^2 m_B^2}  h_-^{(1)} q^2 \nonumber\\
&&+ h_0^{(0)}\sqrt{q^2} + h_0^{(1)}(q^2)^\frac{3}{2}  +{\cal O}((q^2)^\frac{5}{2})\,. 
\eea
Notice that, compared to $h_\pm$, $h_0$ enters the decay amplitude with an additional factor of $\sqrt{q^2}$, which is the reason why we keep only two terms in its expansion. In our fits, the parameters $h_ \lambda^{(i)}$ are allowed to be complex, and we consider the following prior ranges for both their real and imaginary parts:

\begin{table}[!t]
\centering
\renewcommand{\arraystretch}{1.5}
{\footnotesize
\begin{tabular}{|c|c|c|c|}
\hline
\phantom{aa}$\vert\Delta C_{9,2}(q^2)\vert$\phantom{aa} &  \phantom{aa}\emph{data driven}\phantom{aa} & \emph{model dependent} & QCDF \\
\hline
$\vert\Delta C_{9,2}(1.0)\vert$ &   [2.13,   4.37] & [0.54,   1.20] & [0.38, 0.46] \\
\hline
$\vert\Delta C_{9,2}(1.5)\vert$ &   [2.43,   4.21] & [0.35,   0.81] & [0.33, 0.41] \\
\hline
$\vert\Delta C_{9,2}(2.0)\vert$ &   [2.51,   4.04] & [0.25,   0.62] & [0.28, 0.37] \\
\hline
$\vert\Delta C_{9,2}(2.5)\vert$ &   [2.50,   3.84] & [0.19,   0.51] & [0.26, 0.33] \\
\hline
$\vert\Delta C_{9,2}(3.0)\vert$ &   [2.44,   3.64] & [0.16,   0.44] & [0.25, 0.33] \\
\hline
$\vert\Delta C_{9,2}(3.5)\vert$ &   [2.35,   3.43] & [0.13,   0.39] & [0.25, 0.33] \\
\hline
$\vert\Delta C_{9,2}(4.0)\vert$ &   [2.25,   3.22] & [0.12,   0.36] & [0.25, 0.33] \\
\hline
$\vert\Delta C_{9,2}(4.5)\vert$ &   [2.14,   3.04] & [0.11,   0.34] & [0.25, 0.34] \\
\hline
$\vert\Delta C_{9,2}(5.0)\vert$ &   [2.01,   2.88] & [0.11,   0.33] & [0.25, 0.35] \\
\hline
$\vert\Delta C_{9,2}(5.5)\vert$ &   [1.87,   2.74] & [0.11,   0.34] & [0.26, 0.36] \\
\hline
$\vert\Delta C_{9,2}(6.0)\vert$ &   [1.72,   2.64] & [0.11,   0.34] & [0.26, 0.36] \\
\hline
$\vert\Delta C_{9,2}(6.5)\vert$ &   [1.58,   2.55] & [0.11,   0.34] & [0.26, 0.37] \\
\hline
$\vert\Delta C_{9,2}(7.0)\vert$ &   [1.43,   2.50] & [0.20,   0.41] & [0.27, 0.38] \\
\hline
$\vert\Delta C_{9,2}(7.5)\vert$ &   [1.30,   2.46] & [0.34,   0.63] & [0.30, 0.42] \\
\hline
$\vert\Delta C_{9,2}(8.0)\vert$ &   [1.19,   2.44] & [0.55,   0.95] & [0.36, 0.46] \\
\hline
\end{tabular}
}
\caption{\em 68\% HPDI for the hadronic contribution $\vert\Delta C_{9,2}(q^2)\vert$ entering in $B \to K^*$ and $B \to \phi$ transitions at different values of $q^2$, both in the \emph{data driven} and the \emph{model dependent} approaches. In the last column, we also report the expected size of the contributions coming from QCDF.
\label{tab:DC92_comparison}}
\end{table}
\bea \label{eq:h_lambda_range}
h_-^{(0)} &\in& [0,0.1]\,, \ \ \ \ \, 
h_-^{(1)}  \in  [0,4]\,, \qquad 
\ h_-^{(2)}  \in  [0,10^{-4}]\,, \nonumber \\
h_+^{(0)} &\in& [0,0.0003]\,,
h_+^{(1)}  \in  [0,0.0005]\,,
h_+^{(2)}  \in  [0,10^{-4}]\,, \nonumber \\
h_0^{(0)} &\in& [0,0.002]\,, \ \
h_0^{(1)}  \in  [0,0.0004]\,.
\eea
Such ranges have been chosen with the only requirement that increasing them would not alter the results of our fits, and are representative of our current ignorance within the \textit{data driven} approach, where we refrain ourselves from introducing any kind of theory bias other than the choice of the parameterization.

The direct comparison of the fitted results for the hadronic parameters in the two different scenarios is, for several reasons, a non trivial task. Indeed, as explained above, the hadronic contributions are differently parametrized in the two approaches, with no trivial way to directly relate a set of parameters to the other, due in particular to the presence of form factors in Eqs.~\eqref{eq:hla_BtoKst}-\eqref{eq:h_lambda_range}. Moreover, strong correlations as the one shown in Fig.~\ref{fig:hminus} would have to be taken into account, in order to perform a fair comparison among the two scenarios. Finally, it is also important to remember that while in the \emph{model dependent} approach the hadronic parameters are taken real, this is not the case for the \emph{data driven} case where they are allowed to be complex. Nevertheless, it is still possible to circumvent all these issues in order to perform a meaningful comparison among the two approaches, by simply confronting the obtained values for $\vert\Delta C_{9,i}\vert$, in a similar fashion to what we did graphically in Figure 3 of ref.~\cite{Ciuchini:2021smi}. To this end, we report here the values obtained from the fitted value of the hadronic parameters for the three $\vert\Delta C_{9,i}\vert$ in Tables~\ref{tab:DC91_comparison}-\ref{tab:DC93_comparison}, for values of $q^2$ ranging from 1 to 8 GeV$^2$, both in the \emph{data driven} approach and in the \emph{model dependent} one. As a reference, we show also the expected size of the contributions coming from QCDF.
\begin{table}[!t]
\centering
\renewcommand{\arraystretch}{1.5}
{\footnotesize
\begin{tabular}{|c|c|c|c|}
\hline
\phantom{aa}$\vert\Delta C_{9,3}(q^2)\vert$\phantom{aa} &  \phantom{aa}\emph{data driven}\phantom{aa} & \emph{model dependent} & QCDF \\
\hline
$\vert\Delta C_{9,3}(1.0)\vert$ &   [2.36,   5.98] & [0.83,   1.87] & [0.37, 0.50] \\
\hline
$\vert\Delta C_{9,3}(1.5)\vert$ &   [2.88,   5.75] & [0.52,   1.26] & [0.28, 0.42] \\
\hline
$\vert\Delta C_{9,3}(2.0)\vert$ &   [3.08,   5.54] & [0.37,   0.95] & [0.22, 0.35] \\
\hline
$\vert\Delta C_{9,3}(2.5)\vert$ &   [3.09,   5.28] & [0.28,   0.77] & [0.17, 0.31] \\
\hline
$\vert\Delta C_{9,3}(3.0)\vert$ &   [3.02,   4.98] & [0.22,   0.65] & [0.15, 0.28] \\
\hline
$\vert\Delta C_{9,3}(3.5)\vert$ &   [2.90,   4.66] & [0.18,   0.57] & [0.12, 0.26] \\
\hline
$\vert\Delta C_{9,3}(4.0)\vert$ &   [2.75,   4.33] & [0.16,   0.51] & [0.11, 0.26] \\
\hline
$\vert\Delta C_{9,3}(4.5)\vert$ &   [2.57,   4.02] & [0.14,   0.47] & [0.10, 0.26] \\
\hline
$\vert\Delta C_{9,3}(5.0)\vert$ &   [2.36,   3.73] & [0.13,   0.45] & [0.12, 0.26] \\
\hline
$\vert\Delta C_{9,3}(5.5)\vert$ &   [2.12,   3.48] & [0.12,   0.43] & [0.15, 0.26] \\
\hline
$\vert\Delta C_{9,3}(6.0)\vert$ &   [1.85,   3.27] & [0.12,   0.42] & [0.18, 0.29] \\
\hline
$\vert\Delta C_{9,3}(6.5)\vert$ &   [1.58,   3.11] & [0.11,   0.40] & [0.21, 0.32] \\
\hline
$\vert\Delta C_{9,3}(7.0)\vert$ &   [1.33,   2.99] & [0.18,   0.45] & [0.25, 0.35] \\
\hline
$\vert\Delta C_{9,3}(7.5)\vert$ &   [1.14,   2.92] & [0.26,   0.62] & [0.27, 0.37] \\
\hline
$\vert\Delta C_{9,3}(8.0)\vert$ &   [1.02,   2.90] & [0.34,   0.84] & [0.31, 0.40] \\
\hline
\end{tabular}
}
\caption{\em 68\% HPDI for the hadronic contribution $\vert\Delta C_{9,3}(q^2)\vert$ entering in $B \to K^*$ and $B \to \phi$ transitions at different values of $q^2$, both in the \emph{data driven} and the \emph{model dependent} approaches. In the last column, we also report the expected size of the contributions coming from QCDF.
\label{tab:DC93_comparison}}
\end{table}

\begin{table}[!t!]
\centering
\renewcommand{\arraystretch}{1.5}
{\footnotesize
\begin{tabular}{|c|c|c|}
\hline
\phantom{aa}$\vert\Delta C_{9}(q^2)\vert$\phantom{aa} &  \phantom{aa}\emph{data driven}\phantom{aa} & QCDF \\
\hline
$\vert\Delta C_{9}(1.0)\vert$ &   [2.33,   6.06] & [0.09, 0.19]  \\
\hline
$\vert\Delta C_{9}(1.5)\vert$ &   [2.36,   5.97] & [0.10, 0.20]  \\
\hline
$\vert\Delta C_{9}(2.0)\vert$ &   [2.41,   5.88] & [0.10, 0.20]  \\
\hline
$\vert\Delta C_{9}(2.5)\vert$ &   [2.46,   5.79] & [0.11, 0.21]  \\
\hline
$\vert\Delta C_{9}(3.0)\vert$ &   [2.52,   5.70] & [0.12, 0.22]  \\
\hline
$\vert\Delta C_{9}(3.5)\vert$ &   [2.58,   5.63] & [0.12, 0.22]  \\
\hline
$\vert\Delta C_{9}(4.0)\vert$ &   [2.65,   5.57] & [0.13, 0.23]  \\
\hline
$\vert\Delta C_{9}(4.5)\vert$ &   [2.71,   5.54] & [0.13, 0.23]  \\
\hline
$\vert\Delta C_{9}(5.0)\vert$ &   [2.76,   5.54]& [0.14, 0.24]  \\
\hline
$\vert\Delta C_{9}(5.5)\vert$ &   [2.80,   5.57]& [0.14, 0.24]  \\
\hline
$\vert\Delta C_{9}(6.0)\vert$ &   [2.83,   5.65] & [0.15, 0.25]  \\
\hline
$\vert\Delta C_{9}(6.5)\vert$ &   [2.84,   5.76] & [0.16, 0.26]  \\
\hline
$\vert\Delta C_{9}(7.0)\vert$ &   [2.83,   5.91] & [0.16, 0.26]  \\
\hline
$\vert\Delta C_{9}(7.5)\vert$ &   [2.82,   6.10] & [0.17, 0.27]  \\
\hline
$\vert\Delta C_{9}(8.0)\vert$ &   [2.80,   6.33] & [0.18, 0.28]  \\
\hline
\end{tabular}
}
\caption{\em 68\% HPDI for the hadronic contribution $\vert\Delta C_{9}(q^2)\vert$ entering in $B \to K$ transitions at different values of $q^2$in the \emph{data driven} approach. In the last column, we also report the expected size of the contributions coming from QCDF. 
\label{tab:DC9_BtoK}}
\end{table}

Concerning the $B \to K$ mode, for the \textit{model dependent} approach we include only the non-factorizable effects coming from hard-gluon exchanges, being the soft-gluon induced terms subleading as found in ref.~\cite{Khodjamirian:2012rm}, and $\mathcal{O}(10\%)$ of the (already small) ones introduced for the $B \to K^*$ mode and described by eq.~\eqref{eq:DC9_LCSR}. On the other hand, in the \textit{data driven} approach we apply the same rationale used behind eq.~\eqref{eq:hla_BtoKst} and define
\bea\label{eq:hla_BtoK}
h_{B\to K}(q^2) &=&\frac{q^2}{m_B^2} V_{L}(q^2) h_{B\to K}^{(1)} + h_{B\to K}^{(2)} q^4+{\cal O}(q^6)\,.\nnl
\eea

Once again, the parameters $h_{B\to K}^{(i)}$ are allowed to be complex, and in our fits we allow the following prior ranges for both their real and imaginary parts:
\bea \label{eq:h_lambda_range_K}
h_{B\to K}^{(1)} &\in& [0,10]\,, \qquad
h_{B\to K}^{(2)} \in [0,0.0002]\,.
\eea

Also in this case the ranges have been chosen only taking care that they are large enough in order not to affect the results of our fits. The particularly large range for $h_{B\to K}^{(1)}$ is due to its strong correlation to $C_9^{\rm NP}$, see Figure 5 of ref.~\cite{Ciuchini:2021smi}. Similarly to what done for the $B \to K^\ast$ transition, we report in Table.~\ref{tab:DC9_BtoK} the fitted values for $\vert\Delta C_{9}(q^2)\vert$. Since, as we stated above, in the \emph{model dependent} approach we do not include the soft-gluon effects, negligible in this scenario, we report in the table only the fitted values for this hadronic correction in the \emph{data driven} approach, together with the expected size of the contributions coming from QCDF.

\begin{acknowledgments}
\noindent The work of M.F. is supported by the Deutsche Forschungsgemeinschaft (DFG, German Research Foundation) under grant  396021762 - TRR 257, ``Particle Physics Phenomenology after the Higgs Discovery''. The work of M.V. is supported by the Simons Foundation under the Simons Bridge for Postdoctoral Fellowships at SCGP and YITP, award number 815892. The work of A.P. is funded by Volkswagen Foundation within the initiative ``Corona Crisis and Beyond -- Perspectives for Science, Scholarship and Society'', grant number 99091. A.P. was funded by the Roux Institute and the Harold Alfond Foundation.  This work was supported by the Italian Ministry of Research (MIUR) under grant PRIN 20172LNEEZ. This research was supported in part through the Maxwell computational resources operated at DESY, Hamburg, Germany. M.V. wishes to thank KIT for hospitality during completion of this work.
\end{acknowledgments}

\bibliographystyle{JHEP-CONF}
\bibliography{hepbiblio}

\end{document}